\documentclass[11pt,twoside]{article}
\usepackage{fleqn,espcrc2}

\usepackage{graphicx}
\usepackage[figuresright]{rotating}

\newcommand{\AmS}{{\protect\the\textfont2
  A\kern-.1667em\lower.5ex\hbox{M}\kern-.125emS}}

\title{Investigation of Neutrino Properties in Experiments at Nuclear Reactors:
Present Status and Prospects}

\author{L.A. Mikaelyan\address[MCSD]{Institute of General and Nuclear Physics,  \\ 
        Russian Research Centre Kurchatov Institute, Moscow, Russia}%
}
       
\begin{document}

\begin{abstract}
The present status of experiments that are being performed at nuclear 
reactors in order to seek the neutrino masses, mixing, and magnetic moments, whose discovery 
would be a signal of the existence of physics beyond the Standard Model, is considered, along with 
their future prospects.
\end{abstract}

\maketitle

\section*{Introduction}

Presently, searches for neutrino oscillations and a determination of the neutrino mass structure are among the
foremost lines of investigation in experimental particle physics. After a long period of searches for neutrino oscillations 
in short-baseline experiments, explorations at nuclear reactors have entered a new phase. Investigations in these 
realms are now being conducted at ever greater distances, and, for the first time, the masses of the neutrinos and their 
mixing are being investigated precisely in those regions that are suggested by the observations of atmospheric and 
solar neutrinos.

Specifically, we consider the following experiments and projects:

(i) the CHOOZ experiment that had been performed by a collaboration of laboratories from Italy, France, Russia, and the 
United States of America and which had been devoted to searches for antineutrino oscillations at long distances from 
the reactor used (the experiment had been completed in 1999; the results were published in [1]);

(ii) the Palo Verde reactor-based experiment that had been performed by a collaboration of laboratories from the United 
States of America and which had been aimed at long-baseline searches for neutrino oscillations (the measurements 
had been completed by the middle of 2000; the results obtained in the exposures of 1998 and 1999 were published in 
[2]); 

(iii) the project of the Kr2Det experiment that will seek oscillations at long distances from the 
reactor used the project, 
which will pursue oscillations characterized by rather small mixing angles, currently being 
developed for the 
Krasnoyarsk underground laboratory (600 mwe) [3] (it is foreseen that the development of the project 
will have been completed by 2003);

(iv) the KamLAND experiment that is being performed at Kamioke by researchers from Japan and the 
United States of America 
and which is devoted to searches of neutrino oscillations at very long distances [4] (it is 
expected that the first results will have been obtained in 2002).

The results of these laboratory experiments will make it possible to reveal 
the role of electron neutrinos in the anomalies of atmospheric neutrinos, to verify the hypothesis of solar-neutrino 
oscillations, and to establish the mass 
structure of the electron neutrino within the model involving the mixing of three neutrinos.

By searches of oscillations at long distances, one usually implies experiments in which detectors are positioned at distances of about l km from the reactor employed (long-baseline experiments). Experiments where the reactor-to-detector distance is 100 km or more are referred to as very 
long baseline ones. It should be emphasized that it is owing to impressive advances in developing procedures for reactor-antineutrino detection that long-baseline and very long baseline investigations became possible.

An extensive list of references on the problem of neutrino oscillations from the 
studies of Pontecorvo and his colleagues [5], an account of the theory and
of the phenomenology of this phenomenon, and a description of the experiments 
that had been performed prior to 1997 can be found in the review articles [6, 7].

Another line of neutrino investigations at nuclear reactors focuses on attempts 
at observing the neutrino magnetic moment. A discovery of the neutrino magnetic moment at a level of 
$10^{-11}{\mu}_{B}$ (${\mu}_{B}$ is the Bohr magneton) in a laboratory experiment would be of crucial 
importance for particle physics and neutrino astrophysics [8, 9]. In order to explain so "large" a 
value of the neutrino magnetic moment, it would be necessary to introduce, in the theory of weak 
interaction, the right-handed W boson in addition to the left-handed one; moreover, the interaction 
of the neutrino magnetic moment with the magnetic field in the convective zone of the Sun could 
enhance ${\nu}_{e}\rightarrow {\nu}_{{\mu},{\tau}}$ transitions (spin-flavor precession) and lead to 
the emergence of a correlation between the recorded solar-neutrino flux and the magnetic activity of 
the Sun. It should be noted that such a correlation was indeed observed [10].

The experiments that were performed at the reactors in Rovno [11] and in Krasnoyarsk [12] 
(see also the review article of Derbin [13]) yielded the constraint 
${\mu}_{\nu} \le2\times 10^{-10}{\mu}_{B}$, which is still far from the desired region of values. 
In Section 2 we consider the attempts that are being undertaken at present to improve the 
sensitivity of nuclear-reactor experiments to the neutrino magnetic moment. Projects that are 
based on the use of intense artificial sources of neutrinos and antineutrinos and which seem rather 
promising are beyond scope of this article.

\section{SEARCHES OF ANTINEUTRINOS}
\subsection{Reactor Antineutrinos}

A liquid organic scintillator serves as a target for reactor antineutrinos in all experiments that 
are devoted to searches for neutrino oscillations and which are considered here. Antineutrinos are 
recorded by the products of the inverse-beta-decay reaction
\begin{equation}
\bar{{\nu}_{e}}+p \rightarrow e^{+}+n,
\end{equation}
whose threshold is 1.804 MeV. The cross section for reaction ( 1) can be represented as
\begin{equation}
{\sigma}(E)=9.556\times 10^{-44}(886/{\tau}_{n}) 
\end{equation}
$$
\times [(E-{\Delta})^{2}-m^{2}]^{1/2}(E-{\Delta})(1+\delta) ({\rm cm}^{2}), 
$$
where ${\Delta}$ = 1.293 MeV, the incident-antineutrino energy $E$ and the electron mass $m$ are 
expressed in MeV; 
the quantity ${\delta} \ll 1$ takes into account recoil and weak-magnetism effects and the radiative 
correction [14], and ${\tau}_{n}$ is the free-neutron lifetime expressed in seconds.

The positron kinetic energy $T$ in reaction (1) is related to the absorbed-antineutrino energy by 
the equation
\begin{equation}
T \ \approx E\ -\ 1.8\ {\rm MeV.}
\end{equation}

In the majority of the cases, photons arising in positron annihilation are absorbed in the sensitive volume, with the 
result that the recorded positron energy increases by about 1 MeV in relation to than in (3). 
In all experiments, use is made of the method of delayed coincidences between the signals from 
the positron and the neutron. In the CHOOZ and
 the Palo Verde experiment, neutrons are recorded by the photon cascade having the total energy 
of about 8 MeV and 
arising upon neutron capture by gadolinium nuclei that are introduced in the target 
scintillator. Neither the KamLAND 
nor Kr2Det project employs gadolinium $-$ the neutron signal is generated there by 2.2 MeV 
photons accompanying neutron capture in hydrogen.

In order to analyze the results of relevant experiments, it is of crucial importance to know 
the properties of a reactor 
as a source of antineutrinos. Per gigawatt of thermal power, a nuclear reactor generates more 
than $2\times 10^{20}$ 
electron antineutrinos per second, the majority of which originate from the beta decay of 
nuclear fragments produced 
in the reactor core upon the fission of uranium and plutonium isotopes; about a quarter of 
these antineutrinos fall in 
the energy region above the threshold for reaction (1). Since the second half of the 1970s it 
has been known that the 
fragments of different fissile isotopes emit electron antineutrinos having noticeably different 
spectra. For the fission 
of $^{235}$U, $^{239}$Pu and $^{241}$Pu the most precise information about the spectra in the 
region above 1.8 MeV 
was obtained at the Institute Laue-Langevin (ILL, Grenoble) by the method of conversion of the
 beta spectra measured 
for the mixture of fragments [16]; for $^{238}$U, use is made of the calculated value [17]. 
Data that concern the reactor 
power and the current isotope composition of the burning nuclear fuel and which are necessary 
for computing the 
flux and the spectrum of antineutrinos are presented by the reactor personnel.

With the aim of obtaining a reference for normalizing data from the CHOOZ experiment, which had 
already been planned at that time, the collaboration of College de France (Paris), Kurchatov 
Institute (Moscow), and LAPP (Annecy) measured in 1992-1994 the total cross section for 
reaction (1) at a distance of 15 m from the Bugey reactor, whose power is 2.8 GW. The result 
was [18] 
$$
{\sigma}_{expt} = 5.750\times 10^{-43}cm^{2}/{\rm (fission \ event)} 
$$
\begin{equation}
\qquad \qquad \qquad \qquad  \pm 1.4\%, 
\end{equation}
which is in good agreement with the cross section ${\sigma}_{V-A}$ calculated by taking 
the convolution of the equation cross section from relation (2) and the spectrum of reactor 
electron antineutrinos:
$$
{\sigma}_{expt}/{\sigma}_{V-A}= 0.987 \pm 1.4\% (expt.) 
$$
\begin{equation}
\qquad \qquad \qquad \qquad \pm 2.7\%({\sigma}_{V-A}).
\end{equation}

Thus, the cross section measured experimentally is more accurate than the 
computed value and can serve as a metrological reference for cross sections in the absence 
of oscillations. 
It should be noted that the fine features recently revealed in the emission of reactor 
electron antineutrinos above the threshold 
for reaction (1) increase the error in the cross section (4) by about 0.5\% [19]. 
This circumstance was taken into account in determining the parameters of oscillations in the 
CHOOZ experiment.

In the next section of the article, we will consider the spectrum of reactor electron 
antineutrinos below the threshold for reaction (1), because knowledge of this spectrum is 
necessary for performing and interpreting experiments that 
look for the neutrino magnetic moment.

\subsection{Motivation}

First of all, we recall basic relations that are valid in the case of mixing of two neutrino-mass eigenstates 
${\nu}_{1}$ and ${\nu}_{2}$, with the corresponding masses being $m_{1}$ and $m_{2}$. We have
\begin{equation}
{\nu}_{e}={\cos}{\theta}\times {\nu}_{1}+{\sin}{\theta}\times {\nu}_{2}
\end{equation}
In this case, the survival probability P(${\nu}_{e}\rightarrow {\nu}_{e}$) $-$ that is, the 
probability that a neutrino that is produced in the source used will retain its original 
nature at a distance L (m) from the source $-$ is given by
$$
P({\nu}_{e}\rightarrow {\nu}_{e}) = 1-\sin^{2}2{\theta}\sin^{2}\left(\frac{1.27{\Delta}m^{2}L}{E}\right),
$$
\begin{equation}
\qquad
\end{equation}
where $\sin^{2}2{\theta}$ is the mixing parameter, ${\Delta}m^{2}\equiv m^{2}_{2}-m^{2}_{1}$ 
is the mass parameter, and $E$ (MeV) is the neutrino energy.

In experiments, oscillations are sought by a characteristic distortion of the spectrum of 
electron antineutrinos (positrons) and by the reduction of the event-counting rate. For 
reactor electron antineutrinos, the relevant distortions 
of the spectrum and the accompanying reduction of the counting rate are maximal, provided that 
\begin{equation}
{\Delta}m^{2}L \ \approx \ 5 \ {\rm eV}^{2} {\rm m.}
\end{equation}
Relations (7) and (8) demonstrate that, for example, at a distance of I km from the reactor, 
the sensitivity of the experiment being discussed is the highest at 
${\Delta}m^{2} \approx 5\times10^{-3}$ eV$^{2}$ and that it becomes poorer fast as 
${\Delta}m^{2}$ decreases.

In the early 1990s, there appeared motivations for seeking reactor-neutrino oscillations in 
the range ${\Delta}m^{2} = 10^{-2}$-$10^{-3}$ eV$^{2}$, which had not yet been explored by 
that time. The investigation of atmospheric neutrinos with the aid of the Kamiokande II 
and IMB Cherenkov detectors in [20] revealed that the ratio 
of the muon-neutrino flux to the electron-neutrino flux is two-thirds as great as its computed 
counterpart. This discrepancy, known as the atmospheric-neutrino anomaly, could be explained 
under the assumption that intense transitions occur through the 
${\nu}_{\mu} \longleftrightarrow{\nu}_{e}$ channel, through the 
${\nu}_{\mu} \longleftrightarrow {\nu}_{\tau}$ channel, or through both these channels 
simultaneously. As to the mass parameter ${\Delta}m^{2}_{atm}$, it could lay within a wide 
range of values between $10^{-1}$ and $10^{-3}$ eV$^{2}$. At present, observations of 
atmospheric neutrinos have yielded even mote compelling 
grounds to believe that neutrino oscillations do indeed exist. If the SuperKamiokande 
data are analyzed under the assumption that only the 
${\nu}_{\mu} \longleftrightarrow{\nu}_{\tau}$ channel is 
operative, the best description is obtained at the following parameter values [21]:
\begin{equation}
{\Delta}m^{2}_{atm} \approx 3\times 10^{-3} \ {\rm eV}^{2}
\end{equation}
(the most probable value), $\sin^{2}2\theta_{atm} > 0.88.$
It should be emphasized that the SuperKamiokande data do not at all rule out a noticeable 
contribution from the ${\nu}_{e} \longleftrightarrow {\nu}_{\mu}$ channel [22].

For more than three decades, the deficit of solar neutrinos in relation to the computed data 
has been considered as a strong argument in favor of the existence of electron-neutrino 
oscillations. An analysis of solar-neutrino data accumulated by 1998 and the inclusion of 
the solar-matter effect, known as the Mikheev-Smirnov-Wolfenstein (MSW) 
effect made it possible to find, in the ($\sin^{2}2{\theta}, {\Delta}m^{2}$) plane, a few 
regions where parameter values ensure a solution to the solar-neutrino problem [23]. 
According to [24], the situation changed as soon as new data 
appeared after the commissioning of the SuperKamiokande facility in 1998. At present, one 
of the solutions $-$ it is referred to as the Large Mixing Angle (LMA) MSW solution 
$-$ provides the best fit to the observation data. The most 
probable values of the parameters that characterize this solution are
\begin{equation}
{\Delta}m^{2}_{sol}\approx 3\times 10^{-5} \ {\rm eV}^{2}, \ \sin^{2}2{\theta}_{sol}\approx 0.8
\end{equation}

\begin{figure}[htb]
\includegraphics*{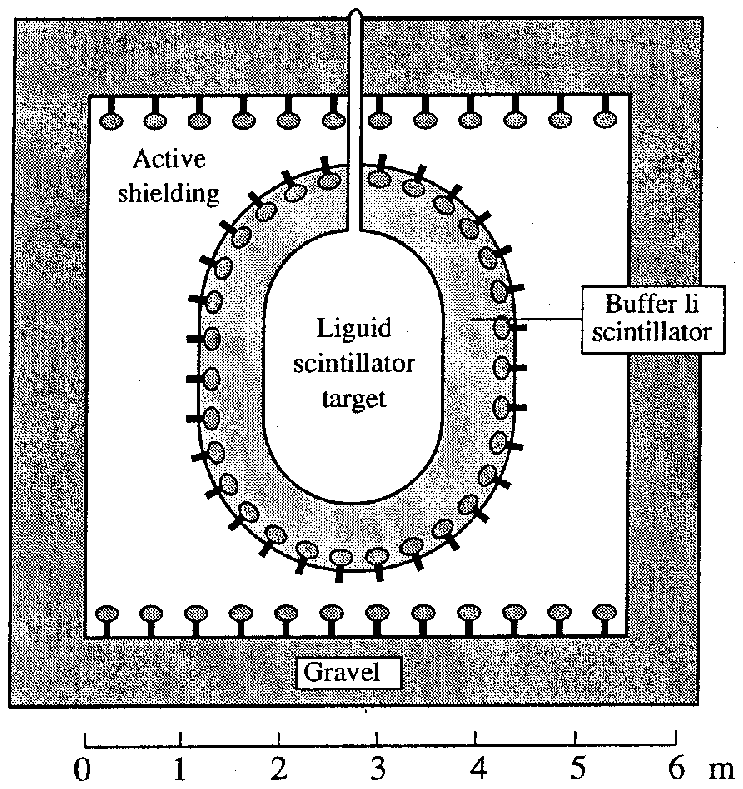}
\caption{Layout of the CHOOZ detector (its description is given in the main body of the text).}
\label{fig:Chooz}
\end{figure}

In the case where two mass eigenstates ${\nu}_{1}$ and ${\nu}_{2}$ are mixed, there is 
obviously one mass parameter ${\Delta}m^{2} = m^{2}_{2} - m^{2}_{1}$. The mixing of at least 
three mass eigenstates is necessary for two mass parameters ${\Delta}m^{2}_{atm}$ and 
${\Delta}m^{2}_{sol}$ to exist. In the case where three mass eigenstates ${\nu}_{1}, {\nu}_{2}$, 
and ${\nu}_{3}$ undergo mixing and where three active neutrino flavors 
${\nu}_{e}$, ${\nu}_{\mu}$ and ${\nu}_{\tau}$ oscillate, there are generally three mass 
parameters: ${\Delta}m^{2}_{21} = m^{2}_{2} - m^{2}_{1}$, 
${\Delta}m^{2}_{31} = m^{2}_{3} - m^{2}_{1}$ and 
${\Delta}m^{2}_{32} = m^{2}_{3} - m^{2}_{2}$. Of these, only two are independent, since 
${\Delta}m^{2}_{21} \equiv {\Delta}m^{2}_{31} - {\Delta}m^{2}_{32}$. According to (9) and 
(10), one of these parameters is two orders of magnitude greater than the other; therefore, 
we have 
\begin{equation}
{\Delta}m^{2}_{sol}= {\Delta}m^{2}_{21} \approx 3\times 10^{-5} \ {\rm eV}^{2}, 
\end{equation}
$$
{\Delta}m^{2}_{atm}\approx {\Delta}m^{2}_{31} \approx {\Delta}m^{2}_{32}\approx 3\times 10^{-3} \ {\rm eV}^{2}.
$$
In the case of reactor neutrinos that is considered here, there are two mixing parameters 
in the adopted scheme. They are expressed in terms of the mixing-matrix elements 
$U_{ei}$ appearing in the superposition 
${\nu}_{e} = U_{e1}{\nu}_{1} + U_{e2}{\nu}_{2} + U_{e3}{\nu}_{3} \quad ({\sum}{U^{2}_{ei}}=1)$; 
that is,
\begin{equation}
\sin^{2}2{\theta}_{LBL}=4U^{2}_{e3}(1- U^{2}_{e3}),
\end{equation}
$$
\sin^{2}2{\theta}_{VLBL}=4U^{2}_{e1}U^{2}_{e2},
$$
where LBL and VLBL stand for, respectively, long-baseline and very long baseline 
reactor-to-detector distances.

Thus, long-baseline and very long baseline reactor experiments make it possible to (i) study 
the role of the electron neutrino in the region of atmospheric-neutrino oscillations; 
(ii) verify whether the hypothesis specified by (10) is 
valid for solar-neutrino oscillations; and (iii) obtain, within the model of three neutrino 
flavors, the full pattern of the 
mass structure of the electron neutrino.

\begin{figure}[htb]
\includegraphics*{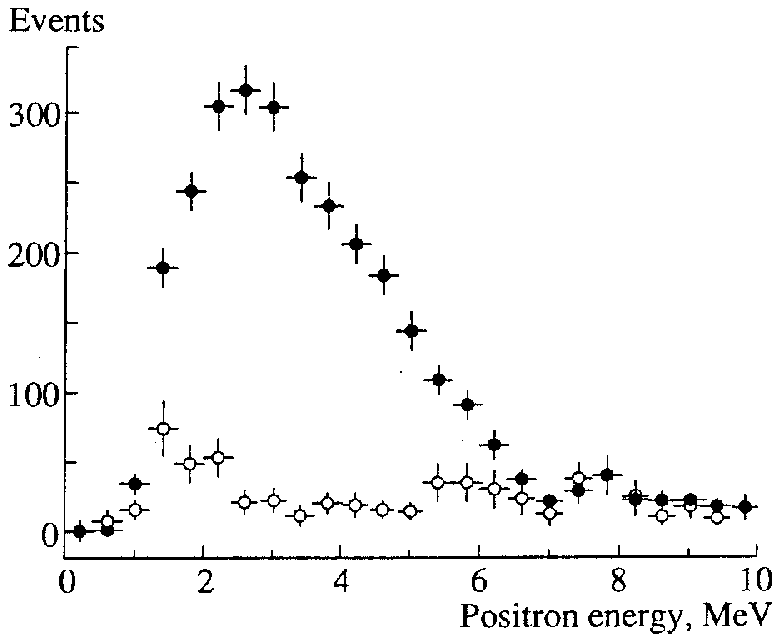}
\caption{Spectra of positrons in the CHOOZ experiment during the reactor (closed circles)
operating and (open circles) shutdown periods.}
\label{fig:spectra}
\end{figure}

The question of the number of neutrino flavors that is greater than three is beyond the scope 
of the present article. 
Yet, it is worth noting that, according to the data of the LSND experiment (Los Alamos), 
${\nu}_{\mu} \rightarrow{\nu}_{e}$ transitions, which are characterized by a rather small 
mixing angle are observed in the region of large mass-parameter values of 
${\Delta}m^{2}_{\rm LSND}\sim $ 1 eV$^{2}$ [25]. The existence of three mass 
parameters, ${\Delta}m^{2}_{sol}$, ${\Delta}m^{2}_{atm}$ and ${\Delta}m^{2}_{\rm LSND}$, 
requires introducing, at least, yet another mass eigenstate ${\nu}_{4}$ and one sterile 
neutrino ${\nu}_{s}$ (either sterile neutrinos do not interact 
with other particle species at all, or the corresponding coupling constant is much less than 
the Fermi constant). The problem of sterile neutrinos has been discussed in the literature 
since the first studies of Pontecorvo [5]; in recent years, interest in this problem has 
become especially acute in connection with reports on the LSND experiments 
(see, for example, [26, 27] and references therein). The potential of nuclear reactors 
for sterile-neutrino searches was schematically considered in [28].

\subsection{Experiments and Projects}
\subsection*{CHOOZ}

The antineutrino detector used was constructed in an underground (300 mwe) gallery at distances 
of 1000 and 1100 m from two PWR reactors. The total rated power of the reactors was 8.5 GW. 
The detector (see Fig. 1) was formed by three concentric spheres. The central zone, which 
contained 5 t of a liquid organic scintillator with an addition of 
gadolinium (about 1 g/l), served as a target for electron neutrinos. The target was 
surrounded by a liquid-scintillator layer (not containing gadolinium) of thickness 70 cm 
followed by third layer (90 t of a scintillator), which acted as a 
passive and an active shielding of the detector. Two inner zones of the detector were viewed by 
192 eight-inch phototubes mounted on a nontransparent screen. The second zone (buffer volume), 
which absorbed annihilation photons and photons arising upon neutron capture in gadolinium 
that escaped from a comparatively small target, 
improved considerably the calorimetric properties of the detector.

The experiment was conducted from April 1997 till July 1998. This was the period within which 
the newly constructed reactors gradually approached the rated mode of operation. This 
circumstance made it possible to have sufficient time for performing measurements during 
the operation of each of the two reactors, with the other reactor being off; during 
the simultaneous operation of the two reactors; and during the period within which the 
reactors were both off (see Table 1).

Neutrino events were required to satisfy the following selection criteria: (a) The energy of 
the first (positron) event and the energy of the second (neutron) event must lay within the 
ranges 1.3$-$8 and 6$-$12 MeV, respectively. (b) The time interval between the positron and 
the neutron event must be in the range 2$-$100 $\mu$s. (c) The spatial conditions are such that 
the distance between the phototube surface and any event must not 
be less than 30 cm and that the first and the second event must not be separated by a distance 
longer than 100 cm. As soon as these selection criteria are imposed, the efficiency of 
neutrino-event detection becomes $\epsilon$ = (69.8 $\pm$1.1)\%.

In all, about 2500 antineutrinos were recorded over the time of measurements, with the measured 
counting rate being 2.58 $\bar{{\nu}_{e}}$/d per 1 GW of reactor power; the typical 
event-to-background ratio was 10 : 1. The positron spectra measured within the reactor 
operating and shutdown periods are displayed in Fig. 2. The ratio $R_{meas/calc}$ of the 
measured neutrino events to that which is expected in the absence of oscillations proved to be 
\begin{equation}
CHOOZ: R_{meas/calc}
\end{equation}
$$
= 1.01\pm 2.8\% (stat.) \pm 2.7\% (syst.).
$$

In this result, the main contribution to the systematic error comes from the uncertainties in 
the reaction cross section (1.9\% $-$ see Subsection 2.1 below), in the efficiency of the 
neutrino-event detection (1.4\%), and in the number of target protons (0.8\%).

\begin{figure}[htb]
\includegraphics*{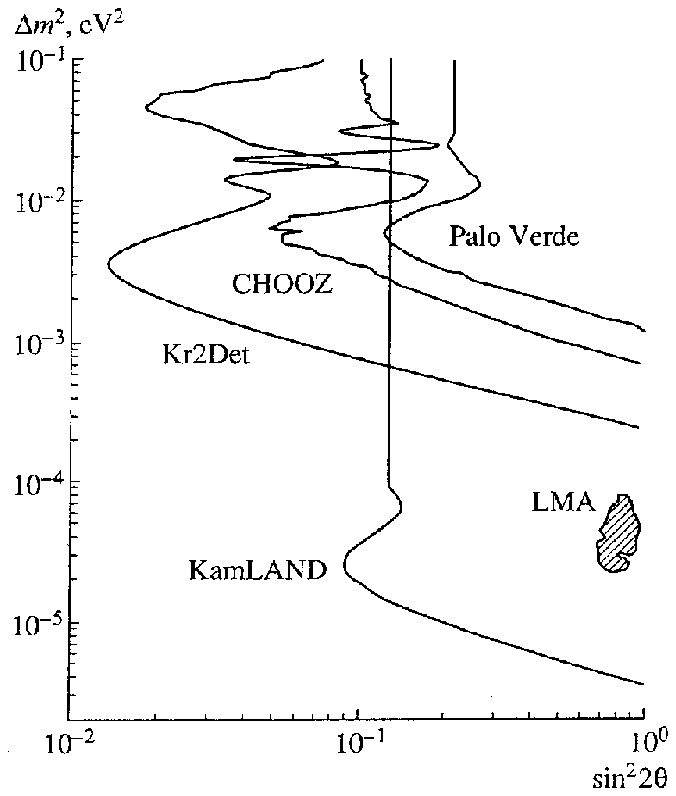}
\caption{Constraints obtained for the oscillation parameters in the CHOOZ and Palo Verde 
experiments at a 90\% C.L. and constraints expected in the Kr2Det and KamLAND projects.}
\label{fig:constr}
\end{figure}

Constraints on the oscillation parameters were obtained by comparing the entire body of 
information accumulated in the experiment with the values that are expected in the absence 
of oscillations. The result (the "CHOOZ" curve in Fig. 3) depends directly on knowledge 
of the absolute values of the characteristics of the neutrino flux and spectrum, 
on the cross section for the inverse-beta-decay reaction, and on the features of the detector. 
As can be seen from Fig. 3, electron neutrinos do not show, at the achieved level of accuracy, 
oscillations in the region ${\Delta}m^2_{atm}$:
\begin{equation}
\sin^{2}2{\theta}_{CHOOZ}\le 0.1, 
\end{equation}
$$
U^{2}_{e3}\le 2.5\times 10^{-2} \ (at {\Delta}m^{2} = 3\times 10^{-3} \ {\rm eV}^{2}).
$$
The result presented in (14) establishes definitively that 
${\nu}_{\mu} \leftrightarrow{\nu}_{e}$ oscillations cannot play a decisive role in the 
atmospheric-neutrino anomaly.

\begin{table}[htb]
\caption{Modes of data accumulation in the CHOOZ experiment}
\label{table:1}
\begin{tabular}{c|c|c|c}
\hline
Reactor 1 & Reactor 2 & Time, d & W, GW\\
\hline
$+$ & 0 & 85.7 & 4.03\\
0 & $+$ & 49.5 & 3.48\\
$+$ & $+$ & 64.3 & 5.72\\
0 & 0 & 142.5 & $-$\\
\hline
\end{tabular}
\end{table}

Searches for oscillations in long-baseline and very long baseline experiments in ever lower 
fluxes of electron antineutrinos require drastically improving the techniques for recording 
reactor neutrinos. The background level of about 0.25 event/d per 1 t of target mass achieved 
in the CHOOZ experiment is nearly 1000 times lower than in any of the previous experiments 
of this kind., In this connection, we would like to note two key points. First, this is the 
underground deployment of the experiment. Under a rock layer of thickness 300 mwe the muon 
flux, which is the main source of a correlated background, decreases in relation to that at 
the Earth's surface by a factor of about 300 down to a level of 0.4/m$^{2}$s. The second 
point is associated with the design of the detector. The introduction of the buffer zone 
(see Fig. 1) reduces the level of the random-coincidence background, shielding the fiducial 
volume from the high natural radioactivity of the phototube glass and structural materials.

\begin{figure}[htb]
\includegraphics*{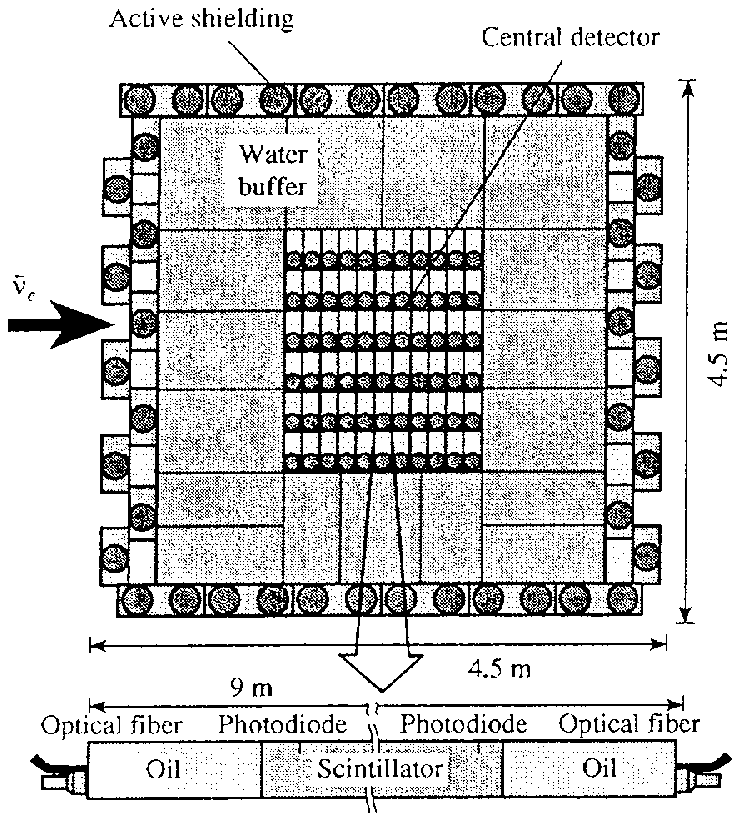}
\caption{Layout of the Palo Verde detector.}
\label{fig:pverde}
\end{figure}

\subsection*{Palo Verde}

In the Palo Verde experiment, three PWR reactors belonging to the same type and having the 
total (thermal) power of 11.6 GW are at distances of 890, 890, and 750 m from a detector 
constructed in an underground laboratory (32 mwe). Once a year, each of the reactors is shut 
down for about 40 days, while the other reactors continues operating over this time interval. 

The experiment being discussed employs a detector whose design is totally different from that 
of the CHOOZ detector and uses more involved methods the for selecting neutrino events. These 
distinctions are dictated by a much more intense muon flux to the detector (about 20/m$^{2}$s) 
and by a less favorable schedule of reactor operation. The electron-antineutrino detector has 
the form of a 6 $\times$ 11 matrix composed of long sections whose dimensions are 
12.7 $\times$ 25 $\times$ 900 cm. The 7.4-m-long central part of each section contains a 
liquid scintillator with an addition of gadolinium; the endface parts of a section are filled 
with mineral oil, each of such endface parts housing a phototube (see Fig. 4). The total 
mass of the target scintillator is 12 t. From all sides, the target is surrounded by 
a purified-water layer (passive shielding) followed by the scintillation sections of an 
active shielding, which generate, in response to the propagation of cosmic-ray muons 
through them, anticoincidence signals, whose frequency is 2 kHz.

A preliminary selection of candidates for events of reaction (1) is accomplished according to 
the following criteria: (a) The positron event must be fast (30 ns), and there must be a 
coincidence between three sections, with the thresholds being 500 keV (positron ionization) 
in one of them and 40 keV (Compton electrons from annihilation photons) in the other two. 
(b) The same conditions must hold for the neutron event. (c) The expectation time for 
the second event must be 450 $\mu$s, which is much longer than the neutron lifetime in 
the scintillator (about 30 $\mu$s); useful and background events are detected in the first 
part of the interval, while the background of random coincidences is recorded within its 
second part. Events satisfying the above criteria were accumulated and were subjected to 
additional amplitude and spatial criteria in the course of a subsequent treatment. 
Upon imposing all the selection criteria, the efficiency effect-to-background ratio of 1 : 1.

The total time of data accumulation in 1998 and 1999 was about 202 days; for 59 days of these, 
two of the three reactors operated. Two different methods were applied to single out the 
neutrino signal. Of these, one was the usual on-off method, which was based on measuring 
the electron-antineutrino flux within the shutdown period of one of the reactors, in which 
case the electron-antineutrino flux from the operating reactors was considered as a 
background component. To the best of my knowledge, the other method was applied for the 
first time. This method made it possible to employ the entire body of accumulated data and 
to separate the useful effect from the background directly in these data. This method, 
dubbed by the authors a swap method, relies on the similarity of the amplitude spectra of 
the first and the second signal in time that stem from a background event and on their 
pronounced distinction in case of the positron and neutron originating from reaction (1).

As a result, it was found that the measured number of neutrino events and that which was 
expected in the absence of neutrino oscillations satisfy the relation 
\begin{equation}
Palo Verde: R_{meas/calc}
\end{equation}
$$
= 1.04\pm 3\% (stat.) \pm 8\% (syst.).
$$

In this result, the systematic uncertainty is presently three times as great 
as the analogous uncertainty in the CHOOZ experiment [see relation (13)]. It is foreseen 
that a further data treatment will lead to a reduction of the uncertainties 
(F. Boehm, private communication). Figure 3 shows the constraints that were obtained for the 
neutrino-oscillation parameters by using the entire body of information accumulated by the 
end of 1999 (Palo Verde curve). 

\subsection*{Kr2Det project}

The Kr2Det project is that of an experiment that will seek oscillations in the region around 
${\Delta}m^{2}_{atm} \sim 3\times 10^{-3}$ eV$^{2}$, but which is expected to have a much higher 
sensitivity to the mixing parameter than the CHOOZ and the Palo Verde experiment. The 
implementation of this project would make it possible to measure the mixing-matrix element 
$U_{e3}$ or to set a more stringent limit on it. It is interesting to note 
that, if the LMA MSW version does indeed solve the solar-neutrino problem, the analysis 
[29] shows that the value of $U^{2}_{e3}$ may be close to the limit that is set by 4
the current constraint $U^{2}_{e3}\le 2.5\times 10^{-2}$.

The basic features of the experiment being discussed are the following:

(i) In order to achieve a higher rate of data accumulation, the target mass is enhanced in 
relation to that of the CHOOZ detector by nearly one order of magnitude. For a target, use 
is made of an organic scintillator of mass 45 t without gadolinium additions.

(ii) In order to eliminate the majority of methodological errors, measurements will be 
performed by simultaneously using two identical spectrometers of electron antineutrinos $-$ 
a far and a near one that are positioned at distances of, respectively, 1100 and 150 m 
from the reactor.

(iii) The experiment will be performed at a depth of 600 mwe, whereby the cosmic-ray component 
of the background is suppressed down to a rather low level.

The detectors have a three-zone structure (see Fig. 5). The phototubes used are mounted on 
a metal sphere that separates zones 2 and 3 by light, which are filled with nonscintillating 
mineral oil. The expected magnitude of a scintillation signal is 100 photoelectrons per 
1 MeV of energy absorbed in the scintillator. The energy resolution is 
${\sigma}\approx 0.14\sqrt{E(MeV)}$.

Candidates for a neutrino event are required to satisfy the following selection criteria: 
(a) The energy of the first (positron) event and the energy of the second (neutron) event 
must lie in the ranges 1.2$-$9.0 and 1.7$-$3.0 MeV, respectively. 
(b) The second event must be recorded within the time window 5$-$500 ${\mu}$s. 
(c) The spatial distances between the events must not exceed 100 cm. The duration of an 
anticoincidence signal is 1000 ${\mu}$s at a repetition frequency of about 10/s. Under these 
conditions, events of reaction (1) are recorded with an efficiency 80\%, the counting rate 
in the far detector being $N_{\nu}=52 \ \bar{{\nu}_{e}}$/d; the expected background level 
is about 10\% of the magnitude of the useful effect.

\begin{figure}[htb]
\includegraphics*{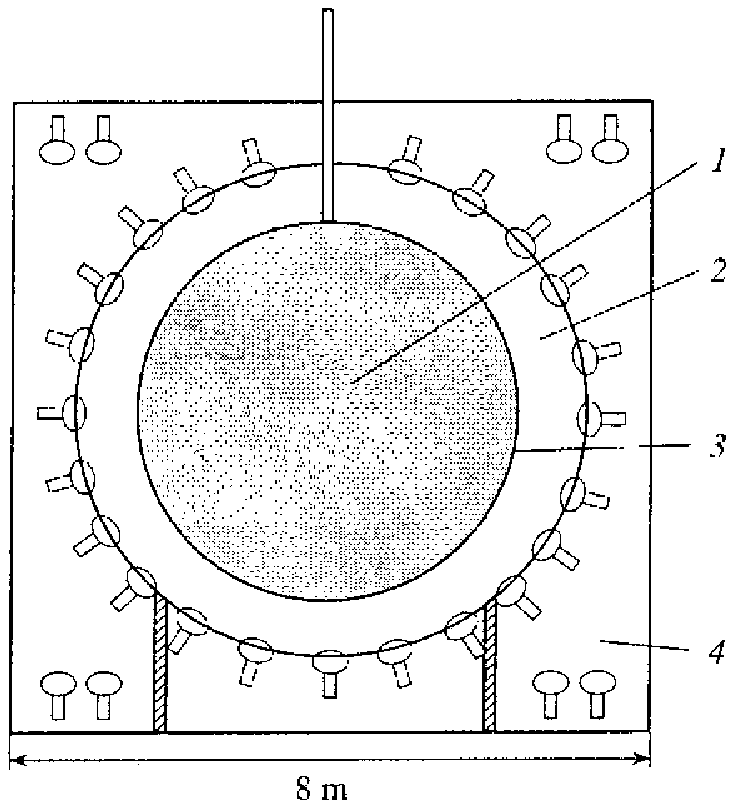}
\caption{Layout of the Kr2Det detector: (1) target (liquid scintillator), (2) buffer (paraffin 
oil), (3) transparent film and (4) anticoincedence region (paraffin oil).}
\label{fig:kr2det}
\end{figure}

In the case where there are no oscillations, the ratio $X_{f,n}$ of the positron spectra 
measured in reaction (1) simultaneously by the far (f) and by the near (n) detectors,
\begin{equation}
X_{f,n} = C\frac{(1 -\sin^{2}2{\theta}\sin^{2}{\phi}_{f})}
{(1 -\sin^{2}2{\theta}\sin^{2}{\phi}_{n})}
\end{equation}
(${\phi}_{f,n}=1.27{\Delta}m^{2}L_{f,n}E^{-1}, \ L_{f,n}$= 1100 and 150 m), is independent of the 
positron energy. Searches for nonzero values of the oscillation parameters 
$\sin^{2}2{\theta}$ and ${\Delta}m^{2}$ are based on 
an analysis of small deviations of the ratio in (16) from a constant value. 
The results of this analysis do not depend on precise knowledge of the spectrum of reactor 
antineutrinos, on the reactor power, on the number of protons in the 
target, or on the detection efficiency. However, the possible distinction between the 
spectral features of the detectors used requires monitoring. A method is being developed 
that would make it possible to reveal this distinction and, if necessary, to introduce 
relevant corrections. It is assumed that the corresponding systematic uncertainty will not 
exceed a few tenths of a percent. The expected cconstraints are displayed in Fig. 3 
(Kr2Det curve). It is foreseen that the experiment will record 40000 electron antineutrinos.

\subsection*{KamLAND}

The electron-antineutrino flux is 1000 times less in this experiment than in the CHOOZ 
experiment. Fifty reactors of an atomic power plant in Japan, which have a total thermal 
power of about 130 GW, serve as a source of antineutrinos. The detector is positioned at 
distances of about 100 to 800 km from the reactors. About 70\% of the electron-antineutrino 
flux is generated by the reactors occurring at distances of 145 to 214 km from the detector. 
The flux never decreases down to zero, but it undergoes seasonal variations, changing by 
$\pm$(10-15)\% around its mean-annual value.

\begin{figure}[htb]
\includegraphics*{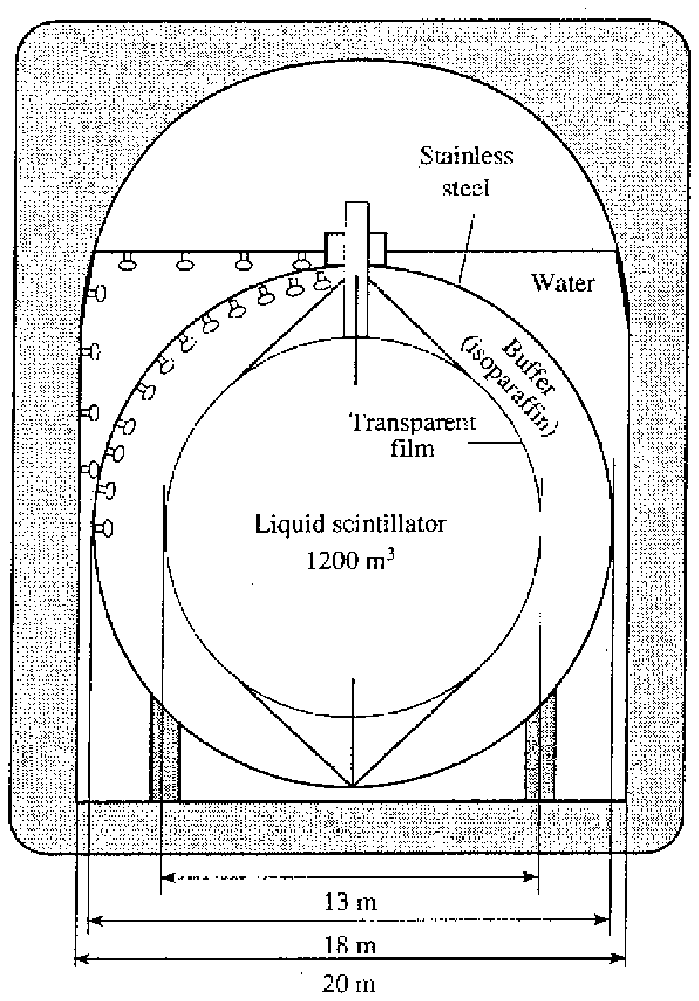}
\caption{Layout of the KamLAND detector.}
\label{fig:kamland}
\end{figure}

\begin{figure}[htb]
\includegraphics*{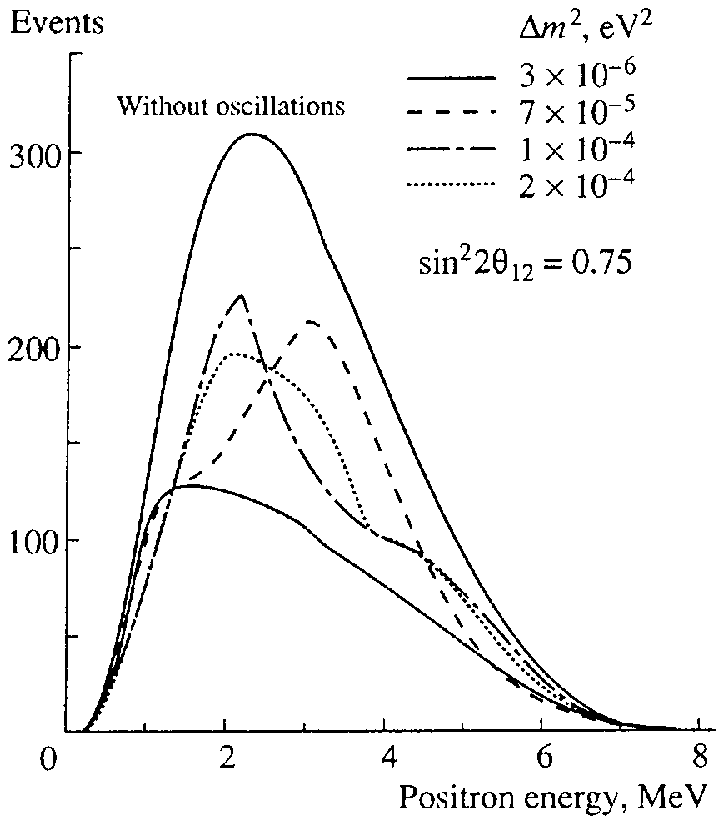}
\caption{Positron spectra in the KamLAND experiment for a few values of the mass parameter
from the LMA MSW region (calculation).}
\label{fig:spkaml}
\end{figure}

The detector is being constructed at a depth of 2700 mwe in the cavern that previously housed 
the Kamiokande facility. Phototubes that cover 30\% of the surface view, through the oil 
layer of the buffer zone of thickness 2.5 m, a spherical target containing 1000 t of a 
liquid scintillator (see Fig. 6). In contrast to the detectors described above, the outer 
layer is filled with water here. In just the same way as in the other cases, this zone plays 
the role of a passive and an active shielding. The expected muon flux to the detector is 
about 0.3/s. In order to reduce the background associated with natural radioactivity, 
it is proposed to purify the target scintillator from uranium an thorium to a level of 
10$^{-16}$g/g. Special measures are taken to prevent the penetration of radon into the 
scintillator. In addition to conventional criteria for selecting neutrino events in amplitude, 
in time, and in positron-neutron spacing, some additional criteria will be imposed to eliminate 
the background (for more details, see [4]). At a 100\% detection efficiency, the neutrino-event 
counting rate computed in the absence of oscillations is about 800/yr, the event-to-background 
ratio being not poorer than 10 : 1. However, it is indicated in the project that in the 
positron-energy region extending up to about 2.5 MeV there must be an irremovable background 
of so-called terrestrial antineutrinos originating from the chains of uranium and thorium 
decays, its magnitude in this energy region being commensurate with the signal from the 
reactors used. In passing, we note that problems associated with studying the antineutrino 
activity of the Earth, which are of prime interest for geology, were posed more than forty 
years ago [30].

In Fig. 7, the positron spectra expected in the experiment are shown for a few values of 
${\Delta}m^2$. It is believed that measurements spanning a period of three years will make 
it possible to establish with confidence whether the electron neutrino oscillates with 
parameters from the LMA MSW region (see the KamLAND curve in Fig. 3).

\section{SEARCHES FOR THE NEUTRINO MAGNETIC MOMENT}

\subsection{Spectra and Cross sections}

Information presented in this section may be of use in considering and planning reactor 
experiments aimed at improving the sensitivity of searches for the neutrino magnetic moment.

\subsection*{Electron-antineutrino scattering on free electrons}

An antineutrino that possesses a magnetic moment ${\mu}_{\nu }$ can be scattered by an 
electron. The cross section for magnetic scattering on a free electron at rest,
$d{\sigma}^{m}/dT$, is proportional to ${\mu}_{\nu }^{2}$ [31]; that is,
\begin{equation}
\frac{d{\sigma}^{m}}{dT}={\pi}r^{2}_{0}\left(\frac{{\mu}_{\nu}}{{\mu}_{B}}\right)^{2}\left(\frac{1}{T}-\frac{1}{E}\right),
\end{equation}
where ${\pi}r^{2}_{0} = 2.495\times 10^{-25}$ {\rm cm}$^{2}$, $E$ is the incident-neutrino 
energy, and ${\mu}_{B}$ is the Bohr magneton.

The cross section for $\bar{{\nu}_{e}}e$ scattering associated with weak interaction (see, 
for example, [8]) is given by
\begin{equation}
\frac{d{\sigma}^{w}}{dT} = G^{2}_{F}\frac{m}{2{\pi}}[4x^{4}+(1 + 2x^{2})^{2}
\end{equation}
$$
\times \left(1-\frac{T}{E}\right)^{2}-2x^{2}(1 + 2x^{2})\frac{mT}{E^{2}}], 
$$
where $x^{2}= \sin^2{\theta}_{W}$ = 0.232 is the Weinberg angle and magnetic interaction and 
$G^{2}_{F}(m/2{\pi})=4.31\times 10^{-45}$ cm$^{2}$/MeV.

\begin{figure}[htb]
\includegraphics*{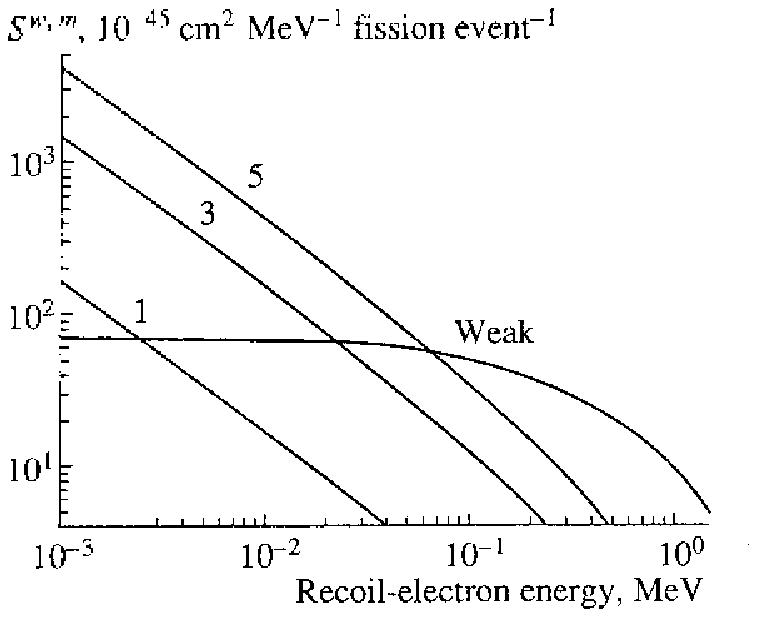}
\caption{Spectra for ${\nu}_ee$ scattering associated with weak and magnetic interaction
(results of the calculations). The figures on the curves representing the spectra for
magnetic scattering are the values of ${\mu}_{\nu}$ in 10$^{-10}{\mu}_{\rm B}$ units.}
\label{fig:magmom}
\end{figure}

For a given value of the incident-neutrino energy, the kinetic energy of the recoil electron 
is constrained by the condition 
\begin{equation}
T\le T_{max} = \frac{2E^{2}}{(2E+m)}. 
\end{equation}

An experiment consists in measuring the total spectrum of recoil electrons upon scattering 
associated with weak and magnetic interaction, $S^{w}(T) + S^{m}(T)$. The spectra $S^{w}(T)$ 
and $S^{m}(T)$ (cm$^{2}$ MeV$^{-1}$ fission event$^{-1}$) are represented as the convolutions 
of the cross sections in (17) and (18) with the reactor-antineutrino spectrum ${\rho}(E)$ 
(MeV$^{-1}$ fission event$^{-1}$). Here, scattering associated with weak interaction $-$ a 
process of importance in its own right-plays the role of a background that is correlated with 
the operation of the reactor used. As the kinetic energy $T$ of the recoil electron decreases, 
the spectrum $S^{m}(T)$ grows indefinitely, whereas the spectrum $S^{w}(T)$ tends to a constant 
limit (see Fig. 8). The two spectra become equal at $T=300 \ (2.5)$ keV for 
${\mu}_{\nu}=10^{-10}{\mu}_{{\rm B}}$ (${\mu}_{\nu}=10^{-11}{\mu}_{{\rm B}}$).

In order to discover the magnetic moment at the level of ${\mu}_{\nu}=10^{-11}{\mu}_{B}$, it 
is therefore necessary to measure the recoil-electron spectra in the energy region below a 
value of about 10 keV. At such low values of the recoil energy, the differential cross sections 
in (17) and (18) for the spectrum of reactor electron antineutrinos assume the asymptotic form
\begin{equation}
\frac{d{\sigma}^{m}}{dT}= 2.495\times 10^{-47} {\rm cm}^{2}/T 
\end{equation}
$$
({\rm for} \quad {\mu}_{\nu}=10^{-11}{\mu}_{{\rm B}}),
$$
$$
\frac{d{\sigma}^{w}}{dT}=10.16\times 10^{-45} {\rm cm}^{2}/{\rm MeV}.
$$

In this approximation, the recoil-electron spectra $S^{m,w}(T)$ are independent of the details 
of the shape of the spectrum ${\rho}(E) \ -$ they are determined exclusively by the total 
number of antineutrinos per fission event, $N_{\nu} = \int{{\rho}(E)dE}$ 
(fission event)$^{-1}$; that is,
\begin{equation}
S^{m}(T) = 2.495\times 10^{-47} N_{\nu}/T 
\end{equation}
$$
{\rm in\ units} \qquad {\rm cm}^{2} MeV^{-1}\ {\rm fission \ event}^{-1}
$$
$$
({\rm for} \ {\mu}_{\nu} = 10^{-11}{\mu}_{{\rm B}}),
$$
$$
S^{w}(T) = 10.16\times 10^{-45} N_{\nu}
$$
$$
{\rm cm}^{2} \ {\rm fission \ event}^{-1}. 
$$

\subsection*{Spectrum of reactor antineutrinos}

The reduction of the recoil-electron-detection threshold increases the contribution that 
electron antineutrinos from the region lying below the threshold for inverse beta decay make 
to scattering associated with magnetic and weak interaction; as was indicated above, nearly 
the entire spectrum of electron antineutrinos comes into play as soon as the threshold becomes 
less than some 15 keV. 

Here, we would like to highlight the qualitative features of the soft section of the 
reactor-antineutrino spectrum that were revealed in recent years [32, 33].

About three-fourths of all electron antineutrinos emitted by a reactor fall within the energy 
range 0$-$2 MeV. A significant contribution to the spectrum in this range comes from 
antineutrinos originating from the beta decay of nuclei that are formed in the reactor core 
upon radiative neutron capture, This contribution can be evaluated on the basis of the data 
given in Fig. 9, which depicts a typical total spectrum of electron antineutrinos and, 
separately, its component associated with the beta decay of fission fragments. 

\begin{figure}[htb]
\includegraphics*{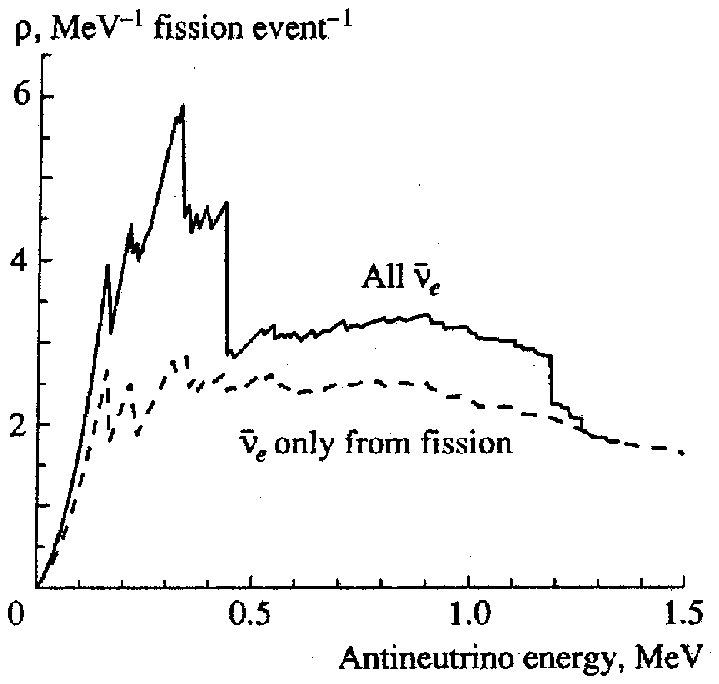}
\caption{Soft section of the spectrum of reactor electron antineutrinos.}
\label{fig:soft}
\end{figure}

The spectrum and the intensity of the electron-antineutrino flux are not determined 
unambiguously by the current reactor state, which is specified by the preset power level 
and by the isotopic composition of the burning nuclear fuel $-$ they also depend on the 
prehistory of this state. From the start of the reactor, there begins a long-term process 
through which the flux approaches its equilibrium value; after the reactor shutdown, 
the flux begins to fall off slowly, and this falloff does not have time to be completed 
by the instant at which the next operating period starts (see Fig. 10). In experiments of 
the type being discussed, the detector background is measured within shutdown periods 
(that is, between two successive operating periods), but, in such periods, there is, in fact, 
a residual radiation of nuclear fuel occurring in the shutdown reactor. The growth of the flux 
and its falloff are accompanied by changes in the spectral content of electron antineutrinos 
and, hence, in the spectra of recoil electrons.

\begin{figure}[htb]
\vspace{9pt}
\includegraphics*{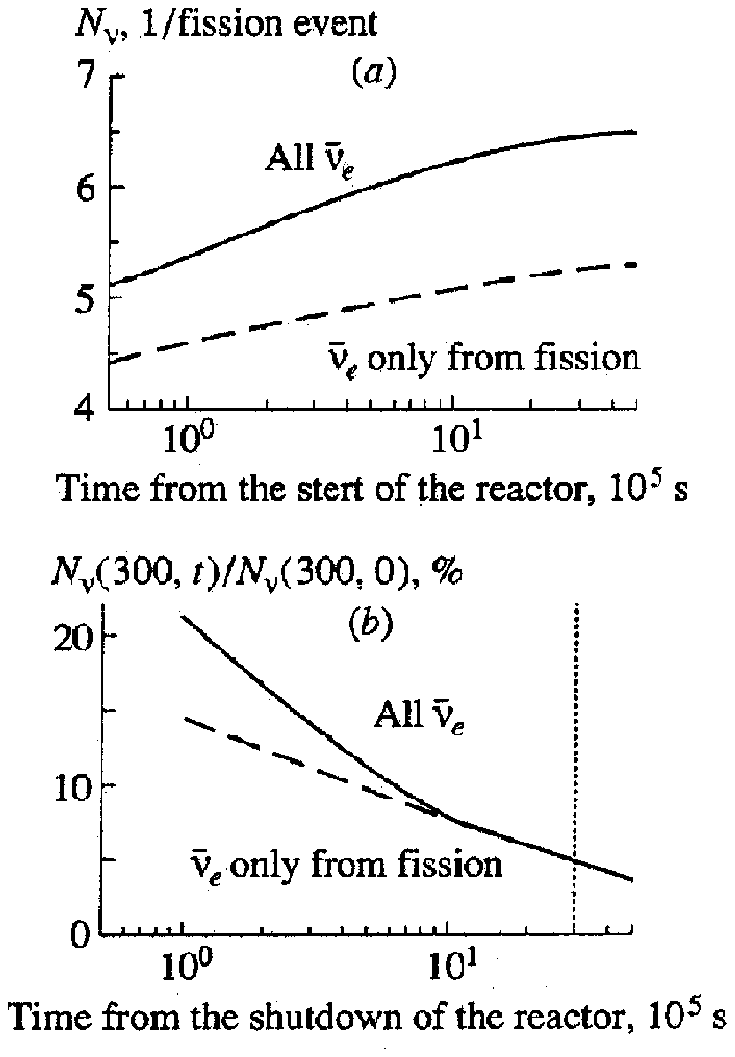}
\caption{Time evolution of the flux of reactor antineutrinos: (a) growth of 
$N_{\nu}=\int{{\rho}(E)dE}$ after the start of the reactor and (b) falloff after the shutdown
of the reactor (in percent with respect to the flux at the end of the reactor operating period).
The vertical dotted line in Fig. 10b indicates the instant at which the next operating period 
begins.}
\label{fig:time}
\end{figure}

In order to describe accurately the total spectrum of reactor antineutrinos, it will be 
necessary to refine further available data on the contributions of its individual components 
and their time dependences and to make use of specific information about the previous 
operation of the reactor used throughout two to three years. As a rule, such information 
can be provided by the personnel of the corresponding atomic power plant. The need for such 
refinements is dictated by the demands of the experiments that are being performed and planned.

\subsection*{Inelastic scattering on atomic electrons}

As the energy lost by a neutrino in a collision event decreases, effects associated with the 
binding of atomic electrons become operative. In the case of inelastic scattering on an electron 
occurring in the $i$th subshell, the energy transfer $q$ from the electron antineutrino is equal 
to the sum of the kinetic energy of the knock-on electron and its binding energy ${\epsilon}_i$ 
in this subshell; that is, 
\begin{equation}
q = {\epsilon}_i + T. 
\end{equation}
The filling of the vacancy formed is accompanied by the emission soft X-ray photons and Auger 
electrons of total energy ${\epsilon}_i$ that is absorbed in a detector material. As a result, 
the event energy observed in the experiment being discussed coincides with the energy transfer 
$q$ in a collision. The differential cross sections and spectra for inelastic scattering on an 
electron of the $i$th shell due to magnetic and weak interaction vanish for $q\le {\epsilon}_i$.

\begin{table*}[htb]
\caption{Binding energies (in keV) of the electrons in the iodine ($Z$=53) and the 
germanium ($Z$=32) atom}
\label{table:2}
\begin{tabular}{c|c|c|c|c|c|c|c|c|cc}
\hline
$Z$ & $\left. 
\begin{tabular}{c}
1$s_{1/2}$\\
$K$\\
\end{tabular}
\right.
$
& $\left. 
\begin{tabular}{c}
2$s_{1/2}$\\
$L_I
$\\
\end{tabular}
\right.
$
& $\left. 
\begin{tabular}{c}
2$p_{1/2}$\\
$L_{II}$\\
\end{tabular}
\right.
$
& $\left. 
\begin{tabular}{c}
2$p_{3/2}$\\
$L_{III}$\\
\end{tabular}
\right.
$
& $\left. 
\begin{tabular}{c}
3$s_{1/2}$\\
$M_I$\\
\end{tabular}
\right.
$
& $\left. 
\begin{tabular}{c}
3$p_{1/2}$\\
$M_{II}$\\
\end{tabular}
\right.
$
& $\left. 
\begin{tabular}{c}
3$p_{3/2}$\\
$M_{III}$\\
\end{tabular}
\right.
$
& $\left. 
\begin{tabular}{c}
3$d_{3/2}$\\
$M_{IV}$\\
\end{tabular}
\right.
$
& $\left. 
\begin{tabular}{c}
3$d_{5/2}$\\
$M_{V}$\\
\end{tabular}
\right.
$ 
 \\
\hline
53 & 32.9 & 5.09 & 4.78 & 4.48 & 1.03 & 0.90 & 0.84 & 0.61 & 0.60\\
32 & 10.9 & 1.37 & 1.22 & 1.19 &  &  &  &  & \\
\hline
\end{tabular}
\end{table*}

To a precision of 2 to 3\%, the results obtained by numerically calculating the spectra for 
the magnetic-interaction-induced ($S^{m}_{in}$) and the weak-interaction-induced ($S^{w}_{in}$) 
inelastic scattering of reactor antineutrinos on the electrons of a iodine (Z = 53) and a 
germanium (Z = 32) atom can be approximated as [34]
\begin{equation}
S^{m,w}(q)\approx \left[\frac{1}{Z}\sum_{i} n_{i}{\theta}(q-{\epsilon}_{i})\right]S^{m,w}_{free}(q),
\end{equation}
where summation is performed over the subshells of the atom involved; $n_i$ is the number of 
electrons in the $i$th subshell; ${\theta}(q-{\epsilon}_{i}$) is the Heaviside step function, 
which is equal to unity for $q \ge {\epsilon}_{i}$ and to zero for $q < {\epsilon}_{i}$; and 
$S^{m,w}_{free}(q)$ is the kinetic-energy spectrum for magnetic interaction-induced 
(weak-interaction-induced) scattering on free electrons (see above), in which case 
${\epsilon}$ = 0 and $q = T$. It is worthy of note that, in this approximation, the binding 
of atomic electrons exerts the same effect on magnetic-interaction-induced and 
weak-interaction-induced scattering.

The calculated binding energies of electrons are given in Table 2.

The actual calculations of inelastic scattering were performed in the energy-transfer ($q$) 
range from 1$-$1.5 to 200$-$300 keV for the $K, L$, and $M$ shells of the iodine atom and the 
$K$ and $L$ shells of the germanium atom, the remaining electrons being considered to be free.

Relation (23) can be formulated in the form of the following rule:

In order to find the distribution of observed energies for the case of inelastic scattering 
on an atom due to magnetic (weak) interaction, it is necessary to compute the spectrum of 
kinetic energies for inelastic scattering on a free electron due to magnetic (weak) interaction 
and multiply the result by the response function $R$,
\begin{equation}
R= \frac{1}{Z}\sum_{i} n_{i}{\theta}(q-{\epsilon}_{i}),
\end{equation}
which depends only on the binding energy of the electrons in the atom. 

As can be seen from relations (23) and (24) and from Fig. 11, the spectra for elastic scattering do not differ from the spectra for inelastic scattering if the energy transfer exceeds the binding energy of a $K$ electron in the target atom. As the energy transfer decreases, an ever greater number of internal atomic electrons successively cease to take part in the scattering process, with the result that the spectra for inelastic scattering constitute an ever smaller fraction of the spectra for elastic scattering. At the energy-transfer value as low as 1 keV, the ratio of the spectrum for inelastic scattering to the spectrum for elastic scattering reduces to 41/53 for iodine and to 22/32 for germanium.

The authors of [34] discussed the accuracy and the applicability range for the recipe in (23) and presented some examples where this recipe is hardly workable or where it is not at all applicable.

\subsection{Experiments}

The experiments reported in [11, 12], as well as the earlier experiment described in [35], 
were intended for verifying theoretical predictions for the structure of weak ${\nu}_{e}e$ 
interaction. In practice, it turned out, however, that the main problem that arises in 
detecting single electrons from $\bar{{\nu}_{e}}e$ scattering is that of the detector background, 
which could not be reliably removed despite massive efforts mounted for many years to solve 
this problem. As a result, it proved to be impossible to test, in reactor experiments, 
the Standard Model in the sector of first-generation leptons, which is the clearest sector 
from the theoretical point of view. The cross section for scattering due to weak interaction 
was experimentally determined in [11, 12] to a relative precision of 50\%. Searches for the 
neutrino magnetic moment involve still greater difficulties.

\begin{figure}[htb]
\includegraphics*{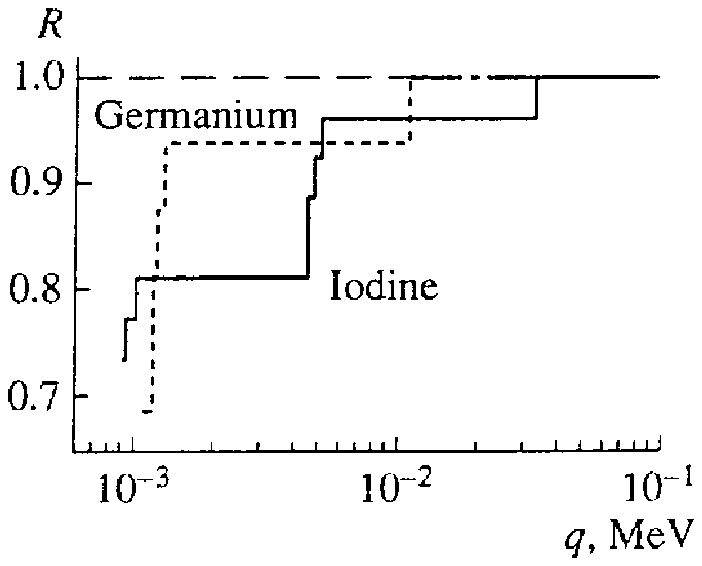}
\caption{Response functions for iodine and germanium [see Eq.(24)].}
\label{fig:respf}
\end{figure}

\begin{table}[htb]
\caption{Number of recoil electrons for a target exposure of 1000 kg d that are associated
with electron-antineutrino scattering on free electrons due to weak (w) and magnetic (m)
${\mu}_{\nu}=3\times 10^{-10}{\mu}_{\rm B}$ interaction}
\label{table:3}
\begin{tabular}{c|c|c|c|c}
\hline
Range of & \multicolumn{2}{c|}{Bugey} & \multicolumn{2}{c}{Krasnoyarsk}\\
recoil-electron & \multicolumn{2}{c|}{ } & \multicolumn{2}{c}{ }\\
\cline{2-5}
energies, keV & w & m & w & m\\
\hline
1$-$4 & 11 & 122 & 6 & 67\\
4$-$16 & 43 & 120 & 24 & 66\\
16$-$60 & 140 & 100 & 77 & 55\\
60$-$250 & 360 & 90 & 200 & 50\\
250$-$1000 & 750 & 50 & 410 & 27\\
\hline
\end{tabular}
\end{table}

In this section, we will consider attempts at reducing the limit on ${\mu}_{\nu}$ that 
are being undertaken by the MUNU collaboration (Grenoble-Munster-Neuchatel-Padova-Zurich) at 
the reactor in Bugey [36] and by the Kurchatov Institute in a collaboration with the 
Petersburg Nuclear Physics Institute at the reactor in Krasnoyarsk [37]. Also, mention 
is briefly made of new-type detectors developed at the Institute for Theoretical and 
Experimental Physics (ITEP, Moscow) and at the Joint Institute for Nuclear Research 
(JINR, Dubna), The expected event-countlng rates that are quoted in Table 3 give an idea 
of the orders of magnitude of the quantities with which one has to deal in experiments 
studying electron-antineutrino scattering on electrons.

\subsection*{MUNU}

The MUNU collaboration has constructed a time-projection chamber (TPC) of volume about 
1 m$^3$, the chamber being filled with a CF$_{4}$ gas. At a pressure of 5 atm the target 
mass is 18 kg. The chamber is surrounded by a liquid scintillator (LS) playing the role of an 
active and a passive shielding (see Fig. 12). The gas circulates, passing through filters 
absorbing oxygen. Since 1998, the detector has been arranged at a distance of 18.6 m from the 
center of the Bugey reactor. The structural materials of the reactor edifice that are situated 
above the detector and which are of thickness approximately equal to 20 mwe ensure shielding 
from the hadronic component of cosmic rays.

\begin{figure}[htb]
\vspace{9pt}
\includegraphics*{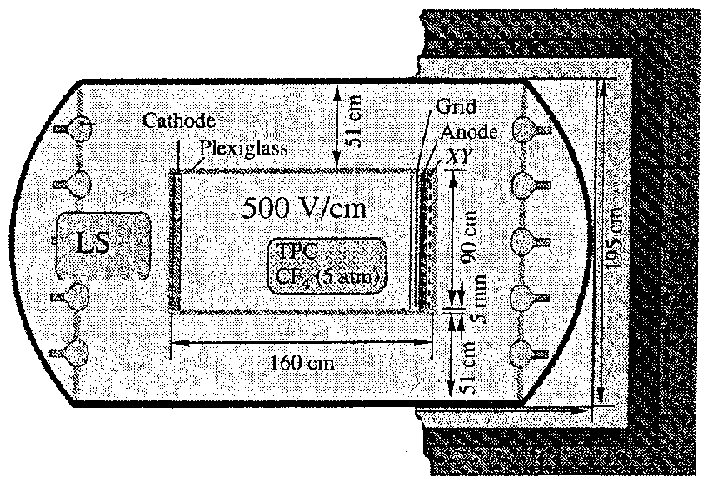}
\caption{MUNU detector of $\bar{{\nu}_e}e$ scattering in Bugey.}
\label{fig:munu}
\end{figure}

The experiment being discussed as measured the energy and the angular distribution of electrons 
with respect to the momentum of incident electron antineutrinos. Within the reactor shutdown 
periods, it was found that the distribution of background events is nearly isotropic. However, 
the absolute value of the background proved to be unacceptably large, and test measurements 
were performed at a detection threshold of 800 to 1000 keV. Investigations made it possible 
to establish the origin of the main background sources. It turned out that the oxygen filter 
is a source of radon and that the material of the chamber cathode contains an admixture that 
emits beta particles with an endpoint energy of about 1.2 MeV, The removal of the filter and 
a replacement of the cathode reduced considerably the background, whereupon the threshold was 
lowered to 300 keV 

\subsection*{Krasnoyarsk}

The detector is being arranged at a depth of 600 mwe In the new laboratory room, where the 
flux of electron antineutrinos is about 40\% less than in the MUNU experiment. The target 
for electron antineutrinos consists of 604 silicon crystal detectors forming a compact 
assembly of four hexahrdral matrices, each containing 151 crystals. An individual detector 
is made in the form of a cylinder of diameter 29 mm and height 100 mm. The total mass of 
silicon is 80 kg, A HPGe detector of volume 116 cm$^3$ is positioned at the center of the 
assembly. The carrying part of the matrices is manufactured from radiation-pure fluoroplastic. 
The target is placed in a cooled chamber with walls of oxygen-free copper. Signals from 
individual crystals are transmitted through stepwise channels to vacuum joints and, further, 
to preamplifiers. The vacuum casing of the detector $-$ it is 64 cm in diameter and 62 cm in 
height and is manufactured from titanium of thickness 4 mm $-$ protects the detector from the 
penetration of radon (see Fig. 13). This casing is followed by a few layers
of a passive and an active shielding, the layer closest to the chamber being made of lead.

It is planned that the detector will have been commissioned at the beginning of 2002. 
The electron detection threshold is presumed to be at a level of 50 keV. The sensitivity to 
be achieved in this experiment will crucially depend on the level of the detector background.

\subsection*{Large xenon chamber (ITEP)}

A time-projection chamber filled with liquid xenon whose total mass will be 750 kg is being 
presently developed at ITEP [38]. A scintillation flash arising in xenon upon antineutrino 
scattering on an electron is fixed by photodetectors specifying the instant at which ionization 
electrons begin to drift along the electric field aligned with the chamber axis. After that, 
the electrons are taken away into the gas phase, where their energy and their X and Y 
coordinates are measured. In the total volume occupied by xenon, the central part containing 
150 kg of it will serve as a target proper for antineutrinos, while the remaining xenon, 
that which surrounds the target, will play the role of a passive and an active shielding. 

The threshold for scattering-event detection is planned to be set at a level not exceeding 
l00 keV. According to the estimate of the authors of the project, a sensitivity in the range 
$(3-5)\times 10^{-11}{\mu}_{B}$ will be achieved in the reactor-electron-antineutrino 
flux of $\bar{{\nu}_{e}}$ $2\times 10^{13}/$cm$^2$ s.

At present, a detector prototype containing 150 kg of xenon is being tested (A.G. Dolgolenko 
private communication).

\subsection*{Toward ultralow energies} 

As the threshold for recoil-electron detection is decreased, events associated with scattering 
due to magnetic interaction are concentrated in ever narrower energy intervals (see Table 3). 
Not only does this localization of magnetic-scattering events reduce the background that is 
generated by events of scattering due to weak interaction and which is correlated with the 
reactor operation, but it also diminishes the relative contribution to these intervals from 
the intrinsic detector background, which is the main obstacle to advances toward the region 
of small magnetic moments. 

Semiconductor ionization germanium detectors make it possible to explore the region of 
energies much lower than those investigated in the Bugey and Krasnoyarsk experiments. By 
using a HPGe crystal of mass about 2 kg at the Gran Sasso laboratory, which is situated at 
a depth of 3200 mwe, the Heidelberg-Moscow collaboration demonstrated the possibility of 
achieving, in the range 11$-$30 keV; the background spectral density at a level of 
0.1/(keV kg d) [39]. To the best of my knowledge, this is the best result in the region of 
low energies.

\begin{figure}[htb]
\vspace{9pt}
\includegraphics*{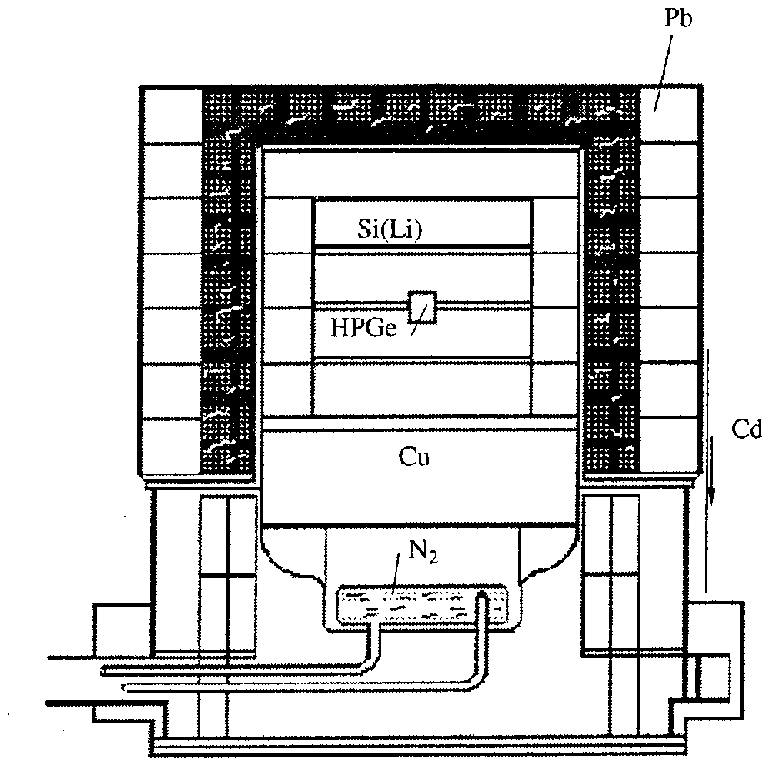}
\caption{Layout of the multicrystal detector for $\bar{{\nu}_e}e$ scattering in Krasnoyarsk
(central part).}
\label{fig:nessy}
\end{figure}

The spectrometer GEMMA, which employs a germanium crystal of mass 2 kg and a system of active 
and passive shielding, has been created and is being tested at ITEP ([40] and A.S. Starostin, 
private communication). It is expected that, at a detection threshold of about 3 keV and a 
20 mwe depth of deployment, the spectrometer background will not exceed 0.3/(keV kg d). 
A cryostat makes it possible to increase the germanium mass up to 6 kg. It is planned that 
the spectrometer will be installed at one of the reactors of the atomic power plant in 
Kalinin. According to the estimate of the authors of the project, the sensitivity to the 
value of ${\mu}_{\nu}$ there will be about $3\times 10^{-11}{\mu}_{B}$ in a flux of 
$2\times 10^{13} \bar{{\nu}_e}$/s over two years of data accumulation.

At present, low-background detectors are being developed for measuring much lower energies 
(see [41] and references therein). These are, first of all, a silicon cryogenic detector 
that employs the effect of ionization-to-heat transition, which was discovered at JINR in 
the 1980s, and, then, a germanium detector involving an internal amplification of ionization 
signals (avalanche germanium detector, also known as AGD ). Employment of such detectors 
will further expand the possibilities for seeking the neutrino magnetic moment. However, 
advances down the scale of electron energies measured in reactor experiments are hampered 
by the emergence of a new form of correlated background, that which is associated with 
recoil nuclei from elastic neutrino-nucleus scattering. Here, at the threshold of the 
absolutely unexplored region, we conclude the section devoted to describing searches for 
the antineutrino magnetic moments in reactor experiments.

\section*{CONCLUSION}

Reactor experiments make it possible to explore the masses of the neutrinos and their mixing 
in the region of small mass parameters not presently accessible to accelerator experiments. 
The CHOOZ experiment established definitively that the ${\nu}_{e}\rightarrow {\nu}_{x}$ 
channel is not dominant in the oscillations of atmospheric neutrinos. KamLAND may become 
the first experiment that will employ terrestrial neutrino sources and which will discover 
the phenomenon of oscillations, determine the contributions of the masses $m_1$ and $m_2$ 
to the electron neutrino, and find a solution to the solar-neutrino problem. The Kr2Det 
experiment, which is characterized by a high sensitivity to small mixing angles, will 
probably be able to reveal the contribution of the mass $m_3$ to the electron neutrino or 
to set a more stringent limit on its value. These investigations rely on unprecedented 
improvements in methods for detecting the inverse-beta-decay reaction.

Searches for the neutrino magnetic moment that are being performed at the reactors in Bugey 
and Krasnoyarsk, an experiment that will employ a large xenon chamber and which is being 
prepared at ITEP, and a foreseen breakthrough into the region of low and ultralow 
recoil-electron energies will presumably permit going beyond the constraint 
${\mu}_{\nu}\le 2\times 10^{-10}{\mu}_{B}$, which could not have been strengthened for 
the three past decades, and expanding the range of searches toward a value of 
${\mu}_{\nu} \sim 10^{-11}{\mu}_{B}$.

The metrological basis of these investigations has become firmer. In particular, 
the features of the flux and of the spectrum of reactor antineutrinos are being refined 
and a simple recipe has been found that makes it possible to determine the cross sections 
for the inelastic scattering of reactor electron antineutrinos on atomic electrons.

\section*{ACKNOWLEDGMENTS}

I am grateful to E. Akhmedov, S. Bilenky, A. Smirnov, P. Vogel, and S. Fayans for valuable 
consultations on the theoretical aspects of this study and to L. Bogdanova, A. Dolgolenko, 
Yu. Kamyshkov, Dy H. Koang, Yu. Kozlov, V. Kopeikin, V. Martem'yanov, A.Pipke, L.Popeko, 
V.Sinev, M. Skorokhvatov, and A. Starostin for discussions on the experiments described in 
this article.

This work was supported by the Russian Foundation for Basic Research (project nos. 
00-02-16035, 00-15-96708).

\end{document}